\documentclass[pra,twocolumn,showpacs]{revtex4}
\usepackage{amssymb}
\usepackage{amsmath}
\usepackage{graphicx}

\begin{document}

\title{
 A 2-level system of cold sodium atoms with high and \\
 tunable susceptibility against the magnetic field}

\author{Z. B. Li}

\author{D. X. Yao}

\author{Y. Z. He}

\author{C. G. Bao}\thanks{Corresponding author: stsbcg@mail.sysu.edu.cn}

\affiliation{The State Key Laboratory of Optoelectronic
Materials and Technologies \\ School of Physics and
Engineering, Sun Yat-Sen University, Guangzhou, P. R. China}

\begin{abstract}
A 2-level spin-system of cold sodium atoms is proposed. Both
the cases that the system has arrived at thermo-equilibrium and
is in the early stage of evolution have been studied. This
system is inert to the magnetic field $B$ in general but very
sensitive in a narrow domain around $B=B_0$, where $B_0$ can be
predicted and is tunable. A characteristic constant
$\gamma=0.278466$ dedicated to various 2-level systems is
found, and leads to a upper limit for the internal energy $U$
of the whole system so that $U\leq\gamma k_B T$. This limit is
considerably lower than the energy assigned to the spatial
motion of only a single particle. Under thermo-equilibrium, the
populations of spin-components measured at distinct $T$
converge to a fixed value when $B=B_0$. The period and
amplitude of population oscillation are found to be seriously
affected by the special sensitivity against $B$. Rich messages
on the dynamic parameters could be obtained via experimental
measurements of these systems.
\end{abstract}

\pacs{
 03.75.Hh,
 03.75.Mn,
 03.75.Nt}

\maketitle

It is widely believed that the systems of cold atoms are
promising in application.\cite{1sten98} Among them the 2-level
systems are notable because the quantum bit might be thereby
defined. In this paper a 2-level system of cold sodium atoms
defined in pure spin-space is studied. This system is inert to
the magnetic field $B$ in general but very sensitive in a
specific domain of $B$. The location of the domain is tunable.
The features of the system under thermo-equilibrium and in the
early stage of evolution
\cite{2chan05,3blac07,4kron05,5kron06,6higb05,7wide05,8wide06,9pech13}
have both been studied. A characteristic constant $\gamma
=0.278466$ is found, and accordingly the internal energy $U$ of
the whole system relative to its ground state (g.s.)
$\leq\gamma k_B T$ disregarding to the particle number $N$ and
the details of dynamic parameters. The above upper limit for
$U$ is considerably smaller than $\frac{3}{2}k_B T$, the energy
assigned to the spatial motion of only a single particle. Thus
the energy involved in the system is extremely low. The
population oscillation is found to be seriously affect by the
special sensitivity against $B$ of the system.

Let $N$ spin-1 sodium atoms be trapped by an isotropic harmonic
potential $\frac{1}{2}m\omega^2 r^2$. When $B$ is not applied,
the Hamiltonian
\begin{equation}
H=\sum_i(-\frac{1}{2}\nabla_i^2+\frac{1}{2}r_i^2)+\sum_{i<j}
\delta(\mathbf{r}_i\mathbf{-r}_j)(c_0+c_2\mathbf{f}_i\cdot
\mathbf{f}_j)  \label{eq1}
\end{equation}
where $\mathbf{f}_i$ is the spin-operator of the $i$-th
particle, $\hbar\omega$ and $\sqrt{\hbar/m\omega}$ have been
used as units for energy and length. We assume that all the
spatial degrees of freedom are frozen (the condition that this
assumption is realized is given below). Note that the
interaction keeps the total spin $S$ and its component $M$
conserved. Thus the eigenstates can be written as
$\Psi_{SM}=\Pi_{i=1}^N\phi_S(\mathbf{r}_i)\vartheta_{SM}^N$,
where $\vartheta_{SM}^N$ is the normalized all-symmetric
spin-state with the good quantum numbers $S$ and $M$. It has
been proved that $\vartheta_{SM}^N$ is unique and $N-S$ must be
even. \cite{10katr01,11bao04} The concrete form of
$\vartheta_{SM}^N$ is irrelevant because related calculations
can be performed by using the fractional parentage
coefficients.\cite{12bao05} Inserting $\Psi_{SM}$ into the
many-body Schr\"{o}dinger equation, the equation for
$\phi_S(\mathbf{r})$ is \cite{11bao04}
\begin{eqnarray}
&&\{\hat{h}_0+[(N-1)c_0+\frac{S(S+1)-2N}{N}c_2]|\phi_S|^2\}\phi_S(\mathbf{r})  \nonumber \\
&=&\varepsilon_S\phi_S(\mathbf{r})  \label{eq2}
\end{eqnarray}
Where $\hat{h}_0=-\frac{1}{2}\nabla^2+\frac{1}{2}r^2$ is the
Hamiltonian of a single particle in the trap. Accordingly, the
energy of $\Psi_{SM}$ is $E_S=\int|\phi_S(\mathbf{r})|^4
d\mathbf{r}(\frac{N(N-1)}{2}c_0+\frac{S(S+1)-2N}{2}c_2)$.

When $B$ is applied, $S$ is no more conserved, but $M$ is. For
a $M$-conserved system the linear Zeeman term is irrelevant,
thus the Hamiltonian arising from the magnetic field is
\begin{equation}
H_B=q\sum_i\mathbf{f}_{iz}^2  \label{eq3}
\end{equation}
where $q=\mu_B B^2/(h^2 E_{HFS})$ and $E_{HFS}$ is the
hyperfine splitting. One can use the set $\Psi_{SM}$ (with a
fixed $M$) to diagonalize $H_B$. The related matrix elements
can be obtained by using the fractional parentage coefficients
as \cite{13bao12}
\begin{equation}
\langle \vartheta_{S'M}^N|H_B|\vartheta_{SM}^N\rangle
=Nq\sum_{\mu}\mu^2 q_{S'S}^{NM\mu}  \label{eq4}
\end{equation}
where
\begin{eqnarray}
q_{S'S}^{NM\mu} &=&\delta_{S'S}[(a_{SM\mu}^{\{N\}})^2+(b_{SM\mu}^{\{N\}})^2]  \nonumber \\
&&+\delta_{S',S-2}a_{S-2,M\mu}^{\{N\}}b_{SM\mu}^{\{N\}}
\nonumber \\
&&+\delta_{S',S+2}a_{SM\mu}^{\{N\}}b_{S+2,M\mu}^{\{N\}}.
\end{eqnarray}
\begin{equation}
a_{SM\mu}^{\{N\}}=C_{1\mu,\ S+1,M-\mu}^{SM}\sqrt{\frac{(N-S)(S+1)}{N(2S+1)}}
\end{equation}
\begin{equation}
b_{SM\mu}^{\{N\}}=C_{1\mu,\ S-1,M-\mu}^{SM}\sqrt{\frac{S(N+S+1)}{N(2S+1)}}
\end{equation}
In the last two equations the Clebsch-Gordan coefficients have
been introduced, $\mu=\pm 1$ or 0 denotes the spin-component of
a particle. After the diagonalization, the eigenstates can be
obtained as $\Psi_{i,M}^B=\sum_S d_S^{B,i}\Psi_{SM}$, where $i$
is a serial number ($i=1$ for the lowest).

We are interested in the systems with $M$ fixed at $N-2$. This
is a 2-level system contains only two basis functions
$\Psi_{N-2,M}$ and $\Psi_{N,M}$. Their spatial wave functions
$\phi_{N-2}(\mathbf{r}_i)$ and $\phi_N(\mathbf{r}_i)$ are very
close to each other
($\langle\phi_{N-2}|\phi_{N}\rangle>0.999$). Therefore we omit
their difference and set $\phi_{N-2}=\phi_N=\phi$. Dropping the
terms that do not depend on $S$, then $E_S=c\frac{S(S+1)}{2}$,
where $c=c_2 \int|\phi(\mathbf{r})|^4 d\mathbf{r}$. The
matrix-equation for diagonalizing $H+H_B$ is 2-dimensional, it
is
\begin{equation}
H_{11}d_1+H_{12}d_2=Ed_1  \label{eq5a}
\end{equation}
\begin{equation}
H_{21}d_1+H_{22}d_2=Ed_2  \label{eq5b}
\end{equation}
Where $d_1\ (d_2)$ is the coefficient of $\Psi_{N-2,M}$
($\Psi_{N,M})$. By using Eq.(\ref{eq4}),
$H_{11}=\frac{c}{2}(N-2)(N-1)+q\frac{2N^2-N-2}{2N-1}$,
$H_{22}=\frac{c}{2}N(N+1)+q\frac{4N^3-4N^2-7N+12}{(2N-1)(2N+3)}$,
$H_{12}=H_{21}=q\frac{2\sqrt{2N-2}}{(2N-1)}$, where $c$ can be
obtained after Eq.(\ref{eq2}) has been solved. The solutions of
Eqs.(\ref{eq5a}) and (\ref{eq5b}) is straight forward. Both the
eigen-energy $E_i$ and the eigen-state
$\Psi_{i,M}^B=d_1^{B,i}\Psi_{N-2,M}+d_2^{B,i}\Psi_{N,M}$ have
analytical forms. In particular, the energy gap $E_{gap}\equiv
E_2-E_1=\sqrt{(H_{11}-H_{22})^2+4H_{12}^2})$. In what follows
the subscript $M$ will be dropped because $M=N-2$ is fixed.

\begin{figure}[tbp]
 \centering \resizebox{0.9\columnwidth}{!}{
 \includegraphics{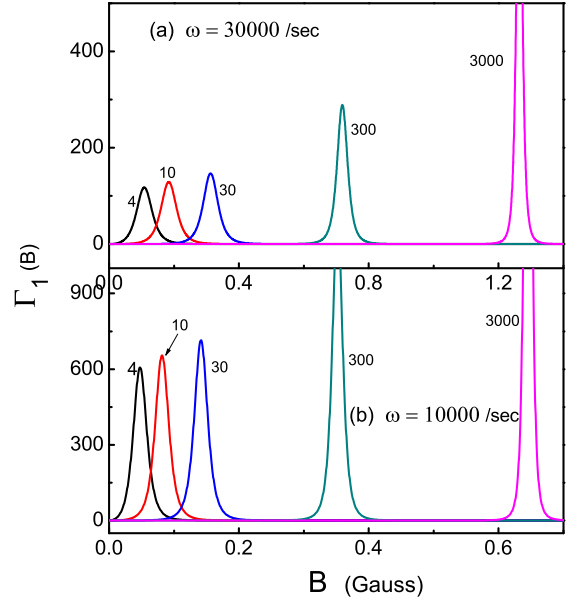} }
 \caption{(color on line) The fidelity susceptibility
$\Gamma_1(B)$ of the g.s. of the 2-level system of Na versus
$B$. $\protect\omega =30000\sec^{-1}$ (a) and
$10000\sec^{-1}$(b), and $N$ is given at five values marked by
the curves. Note that the scales of $B$ are different in (a)
and (b). }
 \label{fig:1}
\end{figure}

We first study the susceptibility of the two eigenstates
against $B$. The fidelity susceptibility is defined as
\cite{14quan06}
\begin{equation}
\Gamma_i(B)\equiv\lim_{\varepsilon\rightarrow 0}
 \frac{2}{\varepsilon^2}(1-|\langle\Psi_i^{B+\varepsilon}|\Psi_i^B\rangle|)  \label{eq6}
\end{equation}
For Na, $\Gamma_i(B)$ of the g.s. distinct in $N$ are plotted
in Fig.1. Due to the fact that $d_1^{B,2}=-d_2^{B,1}$, and
$d_2^{B,2}=d_{1}^{B,1}$, $\Gamma_2(B)=\Gamma_1(B)$, i.e., both
states have exactly the same susceptibility. This is a spacial
feature dedicated only to 2-level systems. There are sharp
peaks in Fig.1 implying that the system is inert to $B$ in
general, but extremely sensitive when $B$ falls in the specific
narrow domains, namely, the domain of sensitivity (D-o-S). The
location of the peak is denoted as $B_{peak}$ which varies with
$N$ and/or $\omega$. The left and right borders of the D-o-S
are named $B_{left}$ and $B_{right}$. They can be roughly
defined as
$\Gamma_i(B_{left})=\Gamma_i(B_{right})=\frac{1}{10}\Gamma_i(B_{peak})$.
A larger $N$ will lead to a higher and narrower peak shifted to
the right, while a larger $\omega$ will lead to a lower and
broader peak also shifted to the right.

\begin{figure}[tbp]
 \centering \resizebox{0.9\columnwidth}{!}{
 \includegraphics{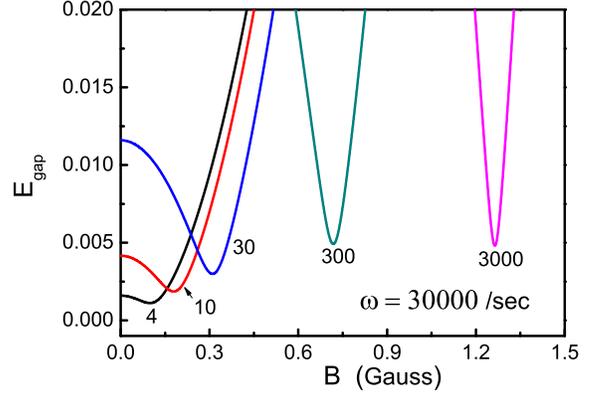} }
 \caption{$E_{gap}$ of the 2-level system versus $B$.
$\omega=30000\sec^{-1}$ is assumed. Refer to Fig.1a. }
 \label{fig:2}
\end{figure}

$E_{gap}$ is plotted in Fig.2. For each $N$, $E_{gap}$ has a
minimum located at $q_0$ ($B_0$). From $\frac{dE_{gap}}{dq}=0$,
\begin{equation}
q_0=\frac{c(8N^2-18)}{4(2N+3)}  \label{eq7}
\end{equation}
(for Na, $q_0$ and $B_0$ fulfill the relation
$q=1745B^2/\omega$, where $q$ is in $\hbar\omega$ and $B$ is in
$Gauss$ \cite{1sten98}). By comparing Fig.2 and 1a, we found
that $B_0\approx B_{peak}$. They are closer to each other when
$N$ is larger (say, $B_{peak}-B_0=0.0096$, 0.0003, and 0 when
$N=4$, $300$, and 3000, respectively). In fact, when we expand
$\Psi_i^{B+\varepsilon }$\ via a perturbation series, we found
that $\Gamma_i(B)\propto (E_{gap})^{-2}$, thus a smaller gap
will lead to a higher sensitivity. Eq. (\ref{eq7}) explain why
a larger $N$ will lead to a larger $B_{peak}$. Furthermore, a
larger $\omega$ will lead to a more compact $\phi(\mathbf{r})$,
and therefore a larger $c$ and a larger $B_{peak}$ as well.
Eq.(\ref{eq7}) provides a convenient way to evaluate the
location of the D-o-S. When $N$ is large, under the
Thomas-Fermi approximation, $c\propto(\omega^2/N)^{3/5}$.
Accordingly, $B_0\propto(\omega^3 N)^{1/5}$. Thus $B_0$
increases with $N$ slowly but increases with $\omega$ rapidly
as shown in Fig.1.

\begin{figure}[tbp]
 \centering \resizebox{0.9\columnwidth}{!}{
 \includegraphics{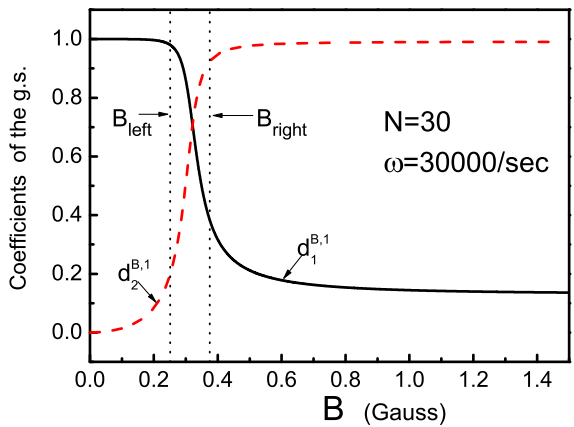} }
 \caption{The coefficients $d_1^{B,1}$ (solid) and $-d_2^{B,1}$
(dash) of the g.s. versus $B$. $N=30$ and $\omega=30000/\sec$
are assumed. }
 \label{fig:3}
\end{figure}

The variation of the g.s. versus $B$ is shown in Fig.3. The
D-o-S appearing in Fig.1a appears again in Fig.3, wherein the
coefficients $d_1^{B,1}$ and $d_2^{B,1}$ undergo a sharp
change. When $B<B_{left}$, $d_1^{B,1}$ remains $\approx 1$
implying that the effect of $B$ is effectively hindered by the
gap. When $B>B_{right}$, we found that
$\Psi_1^B\rightarrow\Pi_{i=1}^N\phi(\mathbf{r}_i)\|N-2,2,0\rangle$.
Where $|N_1,N_0,N_{-1}\rangle$ is a Fock-state in the
spin-space having $N_{\mu}$ particles in $\mu$. Note that, when
$B\rightarrow\infty$, $N_0$ in a g.s. should be maximized
(under the conservation of $M$) to reduce the quadratic Zeeman
energy. Therefore, $\Psi_1^B$ should tend to the above limit.
Whereas the excited state $\Psi_2^B\rightarrow
\Pi_{i=1}^N\phi(\mathbf{r}_i)\|N-1,0,1\rangle$. From Fig.3 we
know that, when $B$ increases, the spin-state of the g.s. is
changed from $\vartheta_{N-2,M}^N$ to $|N-2,2,0\rangle$ and the
change happens essentially inside the D-o-S. Accordingly,
matching the change of the g.s., the excited state is changed
from $\vartheta_{N,M}^N$ to $|N-1,0,1\rangle$.

\begin{figure}[tbp]
 \centering \resizebox{0.9\columnwidth}{!}{
 \includegraphics{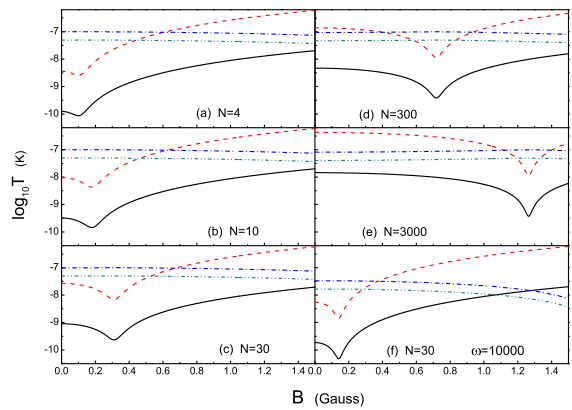} }
 \caption{$T_1$ (solid), $T_2$ (dash), $T_a$ (dash-dot) and
$T_b$ (dash-dot-dot) versus $B$. $\omega=30000\sec^{-1}$(a to
e) and $10000\sec^{-1}$ (f), and $N$ is given at five values
marked in each panel. In the zone below $T_b$ the spatial
excitation can be neglected. In the zone below $T_1$ the system
remains in the g.s.. In the zone between $T_b$ and $T_2$, the
thermo-fluctuation is saturated.}
 \label{fig:4}
\end{figure}

When the system has arrived at thermo-equilibrium under a given
temperature $T$. If $T$ is very low, the spatial excitation is
negligible and the system would be essentially distributed
among the above two eigenstates. Since the energy for spatial
excitation $\approx 1$ (in $\hbar\omega$), the ratio $R_T\equiv
e^{-\beta E_{gap}}/e^{-\beta}$ is crucial, where $\beta=1/(k_B
T)$. We define $T_a$ and $T_b$ at which $R_T=10$ and 100,
respectively. When $T<T_a$, the effect of spatial excitation is
small, when $T<T_b$, the effect of spatial excitation is
negligible. Thus, when $ T<T_b$ the partition function of the
system is simply $Z\approx 1+e^{-\beta E_{gap}}$. The
probability that the system lies at the g.s. is $P_g=1/Z$. We
define further a turning temperature $T_1(B)$\ at which
$P_g=0.95$. When $T<T_1(B)$, not only the spatial but also the
spin degrees of freedom are nearly frozen. Thus $T_1(B)$ marks
the temperature of the secondary
condensation.\cite{15pasq12,16li13} Let $P_{ex}=e^{-\beta
E_{gap}}/Z$ which is the probability lying at the excited
state. Note that when $ T\rightarrow\infty$, an ideal 2-level
system would have $P_{ex}\rightarrow 1/2$. Thus we define the
second turning temperature $T_2(B)$ at which $
P_{ex}=0.95\times 1/2$. When $T=T_2(B)$, the thermo-fluctuation
is close to be saturated. The variations of $T_1(B)$, $T_2(B)$
together with $T_a$ and $T_b$ versus $B$ are plotted in Fig.4.
From the definitions of $T_1(B)$ and $T_2(B)$, one can prove
that they will also arrive at their minimum at $B_0$ as shown
in Fig.4. When $T$ is explicitly lower than $ T_b$, the 2-level
system can be safely considered as a pure spin-system.

The internal energy relative to the g.s. is
\begin{equation}
U(T,B)\equiv E_{gap}e^{-\beta E_{gap}}/Z=E_{gap}P_{ex}  \label{eq8}
\end{equation}
When $T$ is fixed, $\frac{\partial U}{\partial B}$would be zero
if $B=B_0$ and if $e^{-\beta E_{gap}}-\beta E_{gap}+1=0$. For
the latter case, $E_{gap}=1.27846k_B T$ and $U$ arrives at a
maximum $\gamma k_B T$, where $\gamma=0.278466$ is a
characteristic constant dedicated to 2-level systems
disregarding $N$, $\omega$, and the parameters of interaction.
Note that $ \frac{3}{2}k_B T$ \textit{is the energy assigned to
the spatial motion of only a single particle}, while $U$ is the
total spin energy (relative to the g.s.). Thus the very small
upper limit $\gamma k_B T$ manifests how weak the energy is
involved.

\begin{figure}[tbp]
 \centering \resizebox{0.9\columnwidth}{!}{
 \includegraphics{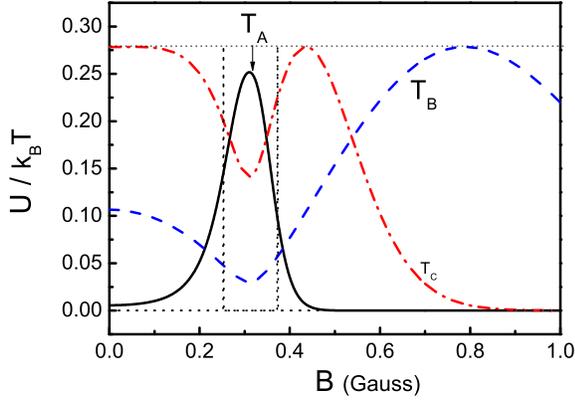} }
 \caption{$U/k_B T$ versus $B$. $N=30$ and $\omega=30000/\sec$
are assumed. The solid, dash, and dash-dot lines are for
$T=T_A$, $T_B$, and $T_C$ (refer to the text), respectively.
The two vertical dotted lines mark the D-o-S, and the
horizontal dotted line marks the upper limit.}
 \label{fig:5}
\end{figure}

An example of $U$ versus $B$ (associated with Fig.4c) are
plotted in Fig.5. $T$ is given at three values,
$T_A=T_1(B_{left})=10^{-9.43}K$,
$T_B=T_2(B_{left})=10^{-7.96}K$, and $T_C=(T_A+T_B)/2$. When
$T=T_A$ and $B$ increases from $B_{left}\rightarrow B_{right}$,
it is clear from Fig.4c that $T_A-T_1(B)$ appears as a small
peak. It implies that $P_{ex}$ undergoes an increase and
afterward a decrease. Accordingly, $U$ is peaked in the D-o-S
as shown by the solid line in Fig.5. Whereas when $T=T_B$ and
$B$ from $B_{left}\rightarrow B_{right}$, the system will enter
to the zone where the thermo-fluctuation is saturated, and
therefore $ P_{ex}$ will remains unchanged. In this case the
variation of $U$ is largely contributed by $E_{gap}$. The dip
in $E_{gap}$ (refer to Fig.2) leads to the dip in $U$ as shown
by the dash curve in Fig.5. When $B$ is larger, the increase of
$E_{gap}$ remains, and thereby $U$ keeps increasing until it
arrives at its upper limit $\gamma k_B T$. When $B$ is larger
further, it is shown in Fig.4c that the system will tend to the
g.s. so that $U$ will tend to zero. It is notable that in all
cases $U\leq\gamma k_B T$ holds as shown by the horizontal
dotted line.

The probability of a particle in $\mu$ in the $\Psi_i^B$ state,
according to Eq.(9) of ref.\cite{13bao12}, is
\begin{eqnarray}
 P_{\mu }^{B,i}
 &=&(d_1^{B,i})^2 q_{N-2,N-2}^{NM\mu}+(d_2^{B,i})^2 q_{N,N}^{NM\mu} \nonumber \\
 &&+2d_1^{B,i}d_2^{B,i}q_{N-2,N}^{NM\mu}
\label{eq9}
\end{eqnarray}

Taking the thermo-fluctuation into account, we define the weighted
probability as
\begin{equation}
\bar{P}_{\mu}^{B,T}=(P_{\mu}^{B,1}+P_{\mu}^{B,2}e^{-\beta
E_{gap}})\//\ Z  \label{eq10}
\end{equation}

\begin{figure}[tbp]
 \centering \resizebox{0.9\columnwidth}{!}{
 \includegraphics{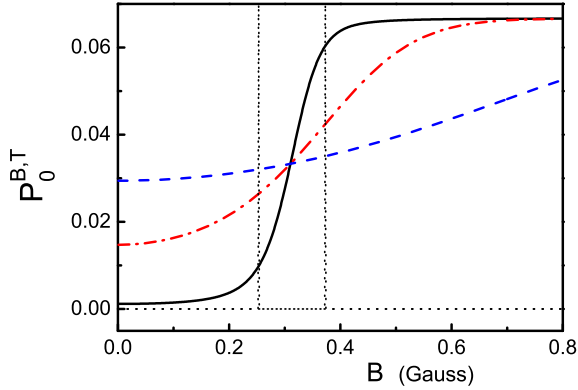} }
 \caption{$\bar{P}_0^B$, the weighted
probability of a particle in $\mu=0$, versus B for the case of
Fig.4c. Refer to Fig.5. }
 \label{fig:6}
\end{figure}

For the case of Fig.4c, $\bar{P}_0^{B,T}$ is plotted in Fig.6,
where $\bar{P}_0^{B,T}$ changes sharply inside the D-o-S when
$T\approx T_A$. Furthermore, it was found that
$P_0^{B_0,1}=P_0^{B_0,2}=1/N$ when $B=B_0$. It arises because
meanwhile the spin-state is equal to
$\frac{1}{\sqrt{2}}(|N-2,2,0\rangle\mp|N-1,0,1\rangle)$, where
$-(+)$ is for the g.s. (excited state). Therefore,
$\bar{P}_0^{B_0,T}=1/N$ disregarding $T$. Accordingly, the
three curves distinct in $T$ converge at $1/N$. Once
$\bar{P}_0^{B,T}$ has been measured at distinct $T$, from the
point of convergency one can know $N$ and $B_0$, the latter is
related to the dynamic parameters. When $B$ is sufficiently
large, from Fig.4, the system might fall into the g.s., and
accordingly $\bar{P}_0^{B,T}\rightarrow 2/N$ as shown in Fig.6.

\begin{figure}[tbp]
 \centering \resizebox{0.9\columnwidth}{!}{
 \includegraphics{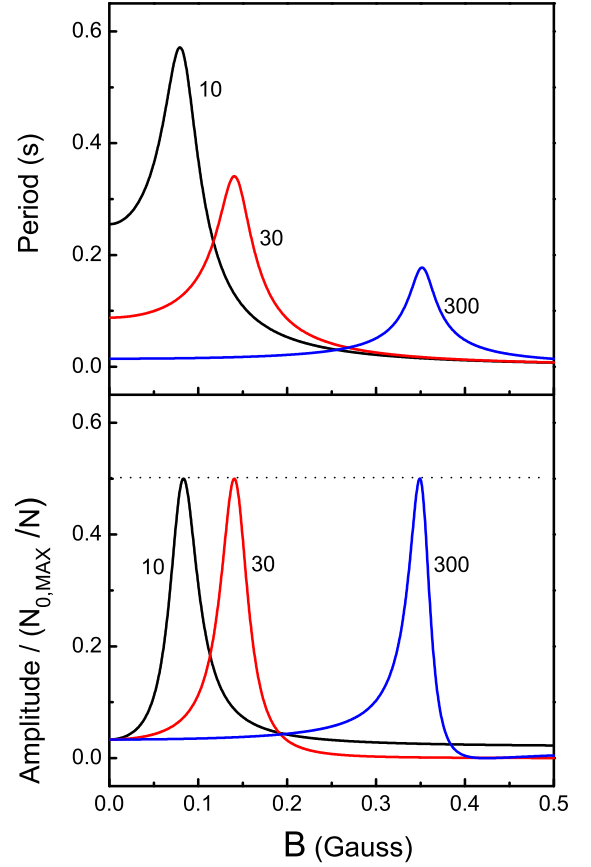} }
 \caption{Period (a) and the amplitude (b) of population
oscillation of Na atoms versus $B$. $\omega=10000$ is assumed
and $N$ is given at three values marked by the curves. The unit
of the amplitude is $N_{0,MAX}/N$, where $N_{0,MAX}=2$ is the
maximal number of $\mu=0$ particles.}
 \label{fig:7}
\end{figure}

\begin{figure}[tbp]
 \centering \resizebox{0.9\columnwidth}{!}{
 \includegraphics{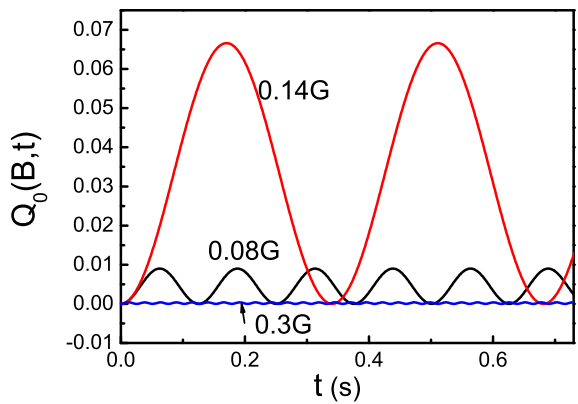} }
 \caption{Population oscillation $Q_0(B,t)$ of Na atoms versus
$t$ (in $\sec$). $N=30$ and $\omega=10000$ are given. $B$ is
given at three values marked by the curves (where, $0.14G=B_0$,
refer to Fig.1b).}
 \label{fig:8}
\end{figure}

There are a number of experiments related to the early stage of
spin-evolution.\cite{2chan05,3blac07,4kron05,5kron06,6higb05,7wide05,8wide06,9pech13}
Starting from an initial state $|\Psi_0\rangle$, the system
will evolve as $\Psi(t)=e^{-iHt/\hbar}|\Psi_0\rangle=\sum_n
e^{-iE_n t/\hbar}|n\rangle\langle n|\Psi_0\rangle$, where $E_n$
and $|n\rangle$ denote the eigenenergy and eigenstate,
respectively.\cite{17chen09} In general there are infinite
eigenstates but only two for our two-level system. Thereby
$\Psi(t)$ has simple analytical form. Let the eigenstate be
re-expanded by the Fock-states as $\Psi_j^B=\Pi_{i=1}^{N}\phi
(\mathbf{r}_i)\ (b_1^{(j)}|N-2,2,0\rangle
+b_2^{(j)}|N-1,0,1\rangle)$, $j=1,2$. Obviously,
$b_1^{(2)}=b_2^{(1)}$ and $b_2^{(2)}=-b_1^{(1)}$, they depend
on $B$. When the initial state is so prepared that
$|\Psi_0\rangle=|N-1,0,1\rangle$ (Namely, $N-1$ particles are
prepared in spin up, then one down-particle is added), the
time-dependent probability of a particle in $\mu=0$ is
\cite{17chen09}
\begin{equation}
Q_0(B,t)=A[1-\cos(E_{gap}\omega t)]  \label{eq11}
\end{equation}
where the amplitude $A=\frac{4}{N}(b_1^{(1)}b_2^{(1)})^2$.
Eq.(\ref{eq11}) provides\ a clear picture of population
oscillation with the period $t_p=\frac{2\pi}{E_{gap}\omega}$.
Examples on $t_p$ and $A$ are given in Fig.7. There are peaks
in both 7a and 7b implying strong oscillation. They match
exactly with the peaks in Fig.1b. However, outside the D-o-S,
the amplitude is very small. When $B=B_0$,
$|b_1^{(1)}|=|b_2^{(1)}|=\frac{1}{\sqrt{2}}$, and therefore
$A=1/N$. and the number of $\mu=0$ particles $N_0(t)\equiv
NQ_0(B,t)$ oscillates from 0 to its maximum $N_{0,MAX}=2$
(refer to the horizontal dotted line in 7b). It is emphasized
that, for the given initial state, $N_0(t)$ can arrives at its
maximum only when $B=B_0$. When $B\rightarrow\infty$,
$E_{gap}\rightarrow 2q\rightarrow\infty$ while
$b_1^{(1)}\rightarrow 1$ and $b_2^{(1)}\rightarrow 0$.
Therefore $t_p\rightarrow 0$ as shown in 7a, and $A\rightarrow
0$ as shown in 7b, and the oscillation damps. The oscillation
of $Q_0(B,t)$ versus $t$ is shown in Fig.8. The fact that a
slight change of $B$ around $B_0$ leads to a great change in
the population oscillation is notable.

\bigskip

In summary, a two-level system of cold sodium atoms has been
studied. The following features are found.

(i) The system is inert to $B$ in general, but very sensitive
in a specific domain (D-o-S), where $\Gamma_i(B)$ appears as a
sharp peak, and $E_{gap}$ appears as a dip. The locations of
the D-o-S can be predicted and can be tuned by changing $N$
and/or $\omega$.

(ii) When $T$ is sufficiently low, the system is free from the
interference of spatial excitation. There is a characteristic
constant $\gamma=0.278466$ dedicated to 2-level systems.
Accordingly, the upper limit of the internal energy $U$
(relative to the g.s.) is $\gamma k_B T$. It implies that the
$U$ of the whole $N$-body system is even $<\frac{1}{2}k_B T$,
the energy assigned to a single spatial degrees of freedom.

(iii) When $B=B_0$ (where the dip of $E_{gap}$ locates at),
$\bar{P}_0^{B,T}$ distinct in $T$ converge at $1/N$. Once
$\bar{P}_0^{B,T}$ can be measured, the messages on $N$, $B_0$,
and thereby the dynamic parameters can be obtained.

(iv) The spin-evolution depends strongly on $B$. When
$B<B_{left}$, the amplitude of oscillation is small (Fig.7).
When $B>B_{right}$, the oscillation damps due to the sustained
increasing of the $E_{gap}$. When $B$ falls into the D-o-S, a
strong oscillation emerges. Thereby valuable message on the
dynamic parameters can be extracted.

(v) We have found a number of distinguished features for Na.
The crucial point is the existence of the minimum in $E_{gap}$.
Note that $\langle N-1,0,1|\vartheta_{N-2,N-2}^N\rangle=0.9915$
while $\langle N-1,0,1|\vartheta_{N,N-2}^N\rangle =0.1302$.
Therefore, when $B$ is small, the g.s. dominated by
$\vartheta_{N-2,N-2}^N$ contains more $\mu\neq 0$ particles
than the excited state contains. Thus the increase of Zeeman
energy in the g.s. is faster than that in the excited state.
This leads to a decline of the gap. Such a decline will be
ended at $B=B_0$ at which $N_0$ of both states are the same.
Whereas, for Rb, the g.s. contains fewer $\mu\neq 0$ particles.
Thus the decline of the gap in accord with the increase of $B$
does not happen. Accordingly, the minimum of the gap is located
at $B=0$. Hence, the 2-level systems of Rb would have
completely different features.

In this paper we have not considered the interference of
spatial excitation. In a recent paper it is reported that some
features of a gas of Na atoms with multi-spatial-mode can be
explained based on a theory independent of the spatial degrees
of freedom.\cite{9pech13} Similar features between Fig.2 of
that paper and Fig.7 of our paper are found. Nonetheless, when
$T\gtrsim T_b$, how serious the spatial-mode would affect the
spin-motion of our 2-level system deserves to be further
studied.

\begin{acknowledgments}
The project is supported by the
National Basic Research Program of China (2008AA03A314,
2012CB821400,  2013CB933601), NSFC projects (11274393,
11074310, 11275279), RFDPHE of China (20110171110026) and
NCET-11-0547.
\end{acknowledgments}

\end{document}